\documentclass[prl,twocolumn,amsmath,amssymb,superscriptaddress]{revtex4}

\usepackage{graphicx,epstopdf,bm}
\usepackage{soul}
\usepackage{hyperref}
\hypersetup{colorlinks=true,linkcolor=black,citecolor=black,urlcolor=black}
\usepackage[normalem]{ulem}
\usepackage{color} 

\usepackage{amsmath}
\usepackage[export]{adjustbox}
 
\usepackage{sidecap}
\sidecaptionvpos{figure}{c}  

\newcommand{\Fkt}[1]{\,\mathsf {#1}}
\ifx\Tr\renewcommand{\Tr}{\Fkt{Tr}}
\else\newcommand{\Tr}{\Fkt{Tr}}
\fi

\begin{document}

\title{Cavity-Controlled High Harmonic Generation}

%
%
%

 

\author{Zohar Amitay}
\email{amitayz@technion.ac.il}
\affiliation{Schulich Faculty of Chemistry and Institute of Advanced Studies in Theoretical Chemistry, 
Technion-Israel Institute of Technology, Haifa 32000, Israel}

\author{Nimrod Moiseyev}
\email{nimrod@technion.ac.il}
\homepage{https://nhqm.net.technion.ac.il/}
\affiliation{Schulich Faculty of Chemistry and Institute of Advanced Studies in Theoretical Chemistry, 
Technion-Israel Institute of Technology, Haifa 32000, Israel}
\affiliation{Faculty of Physics, Technion-Israel Institute of Technology, Haifa 32000, Israel}

 

\begin{abstract}


Employing non-Hermitian Floquet theory,
the strong-field process of high harmonic (HH) generation 
by a classical continuous-wave field irradiating ground-state atom 
is discovered to be controllable by 
placing the irradiated atom 
inside 
a single-mode quantum cavity initiated even with a single photon.
%
%
Judicious cavity coupling of cavity-free photo-induced   
atomic Floquet states forms polaritonic Floquet states 
that generate side harmonics around the 
(standard no-cavity) odd harmonics.
The different possible cavity-controlled HH spectra,
including also the ones resulting from several cavities in a row,
enable attosecond-pulse sequences different from the one produced 
without a cavity.
Moreover, the present study sets the framework and opens the way for further cavity control over the HH generation process as well as over other strong-field processes






\end{abstract}



\maketitle 


As recognized in the 2023 Nobel Prize~\cite{Nobel2023_Physics_A,Nobel2023_Physics_B},
the celebrated process of high harmonic 
generation (HHG)~\cite{HHG_paper1_Ferray_1988,
HHG_paper2_LHuillier_1991,
HHG_paper3_Kulander_1992,
HHG_paper4_Corkum_1994,
HHG_paper5_Lewenstein_PRA_1994,
Nimrod_Weinhold_PRL_1997,
paper1_attosec_pulse_gen_1996,
paper2_attosec_pulse_gen_1997,
paper3_attosec_pulse_gen_1998,
paper4_attosec_pulse_gen_exp_2001} 
by a strong classical field 
is a nonlinear strong-field optical process of high scientific interest and importance,
both on its own and as a basis for attosecond pulse generation and XUV spectroscopy.
In the basic continuous-wave (cw) HHG process with atoms, the irradiation of a ground-state atom with a linearly-polarized strong cw field of frequency $\omega_{0}$ 
photo-induces the emission of coherent radiation with frequencies that are high harmonics of $\omega_{0}$. 
With an infrared (IR) driving field, 
the emitted radiation can  
extend into the extreme ultraviolet (XUV) region and x-ray region.
A key characteristic and limitation of this HHG process is the emission of high-harmonic (HH) 
radiation only at odd harmonics of $\omega_{0}$.

Here, employing non-Hermotian Floquet theory, 
we discover the process of HHG by a cw classical field  
irradiating ground-state atoms  
to be controllable by placing the irradiated atom 
inside a single-mode quantum cavity that is initiated even with a single photon.
Judicious cavity coupling of cavity-free 
atomic Floquet states (solutions),
which are photo-induced by the strong classical cw field,
forms polaritonic Floquet states 
that, once populated, 
generate side harmonics around the 
(standard no-cavity) odd harmonics.
The specific HH spectrum that is generated depends 
on the combined properties of the whole 
atom-cw field-cavity system, 
i.e., on the specific atomic system, 
the frequency and intensity of the irradiating cw field, 
and the cavity frequency and cavity-atom coupling strength.
The different possible cavity-controlled HH spectra,
including also the ones resulting from several cavities in a row,
enable the production of attosecond-pulse sequences that are different from the one produced without a cavity.
%


To set the theoretical framework, we start
with a cavity-free high harmonic generation,  
where a 
Xe atom,
chosen as the present atomic model system,
is irradiated by a strong classical 
linearly-polarized continuous-wave (cw) field 
of amplitude $\varepsilon_{0}$ and frequency $\omega_{0}$.
The field propagates along the $z$ direction and 
its linear polarization is in the $x$ direction.
In the length gauge and under the dipole approximation, 
the time-dependent Hamiltonian of the irradiated atom 
is given by
\begin{equation}
\hat{H}_{}(x,t) = \hat{H}_{atom}(x) + \hat{d} \; \varepsilon_{0} cos(\omega_{0} t) \, ,
\label{eq:H}
\end{equation}
where $\hat{H}_{atom}$($x$) is the field-free Hamiltonian of the Xe atom, 
which is taken here as a one-dimensional single-active-electron model Hamiltonian
of coordinate $x$~\cite{Nimrod_Aver_PRA_2008} 
(along the direction of the field's polarization), 
and $\hat{d}$=$-e x$ is the corresponding atomic dipole-moment operator 
($-e$ is the electron charge).

Within the non-Hermitian quantum Floquet theory~\cite{NHQM-BOOK}, 
%
%
following complex scaling of the atomic coordinate $x$, 
the time-dependent solution of the 
time-dependent Schr$\ddot{\text{o}}$dinger equation 
of the irradiated atom 
with the 
complex-scaled non-Hermitian Hamiltonian is generally given by
%
\begin{equation}
\left| \Psi_{}(x,t) \right) = 
\sum_{\alpha} a_{\alpha} \left| \chi_{\alpha}(x,t) \right) \, ,
\end{equation}
with 
\begin{equation}
\left| \chi_{\alpha}(x,t) \right) = 
\exp(-i \epsilon_{\alpha} t/\hbar) \left| \Phi_{\alpha}(x,t) \right) \, ,
\end{equation}
%
%
where $a_{\alpha}$, 
$\left| \Phi_{\alpha}(x,t) \right)$
and $\epsilon_{\alpha} = E_{\alpha} - i\Gamma_{\alpha}/2$ are, respectively, the amplitude, 
eigenfunction and complex eigenenergy (quasienergy) 
of a resonance Floquet eigenstate $\alpha$. 
The $E_{\alpha}$ and $-\Gamma_{\alpha}/2$ are, respectively, the real and imaginary parts of the quasienergy $\epsilon_{\alpha}$, 
with $E_{\alpha}$ and $\Gamma_{\alpha}$ being, respectively,  
the resonance energetic position and width.
The amplitudes 
$a_{\alpha}$ are determined by the initial state of the system at $t$=0. 
The Floquet eigenstate $\alpha$ is an eigenstate of the complex-scaled 
non-Hermitian Floquet Hamiltonian $\hat{H}_{f}(x_{\theta},t)$ defined as 
\begin{equation}
\hat{H}_{f}(x_{\theta},t) = -i \hbar \frac{\partial}{\partial t} + 
\hat{H}_{atom}(x_{\theta}) + \hat{d}_{\theta} \; \varepsilon_{0} cos(\omega_{0} t) \, ,
\label{eq:H_f}
\end{equation}
such that
\begin{equation}
\hat{H}_{f}(x_{\theta},t) \left| \Phi^\theta_{\alpha}(x,t) \right) = (E_{\alpha} - i\Gamma_{\alpha}/2) \left| \Phi^\theta_{\alpha}(x,t) \right) \, ,
\end{equation}
where $x_{\theta}$=$x\exp(i\theta)$ 
is the complex-scaled atomic coordinate replacing $x$, 
and $\hat{d}_{\theta}$=$-e x_{\theta}$ is the complex-scaled dipole operator,
with $\theta$ being the complex-scaling parameter set to a proper value~\cite{NHQM-BOOK}.
%
%
It is worth noting that  
each Floquet eigenfunction $\left| \Phi^\theta_{\alpha}(x,t) \right)$ 
satisfies outgoing spatial boundary conditions, 
is square integrable with a normalization to one under $c$-product, 
and 
is orthogonal under $c$-product to all the other Floquet eigenfunctions~\cite{NHQM-BOOK}. For simplicity, from now on we omit the symbol $\theta$. 
%

The Floquet eigenfunctions 
are time periodic 
with the cw-field optical period 
$T_{0} = 2\pi/\omega_{0}$, 
and each of them is thus given by 
\begin{equation}
\left| \Phi_{\alpha}(x,t) \right) = 
\left| \Phi_{\alpha}(x,t+T_0) \right) = 
\sum_{n=-\infty}^{\infty} \exp(i n \omega_{0} t) 
\left| \varphi_{n,\alpha}(x) \right) . 
\end{equation}
where the Fourier component $\left|\varphi_{n,\alpha}(x)\right)$ 
is referred to as the $n$-th channel function of Floquet eigenstate $\alpha$.
%
Since $\left|\varphi_{n,\alpha}(x)\right)$  can be expanded in the basis set of the field-free atomic eigenstates,
$\left| \Phi_{\alpha}(x,t) \right)$ is therefore 
a complicated time-dependent coherent superposition of them.

With the present Floquet Hamiltonian of Eq.~(\ref{eq:H_f}), 
another important characteristic 
existing here  
for a given 
Floquet eigenfunction
is its spatio-temporal dynamical 
symmetry~\cite{NHQM-BOOK,Nimrod_Lein_JPhysChemA_2003},
%
\begin{equation}
\left| \Phi_{\alpha}(x,t) \right) = \text{S}_{\alpha} \, \left| \Phi_{\alpha}(-x,t+T_0/2) \right) 
\text{ with } \, \text{S}_{\alpha}=(+) \text{ or } (-) .
\label{Eq:Floqet_state_symmetry}
\end{equation}
%
This implies that all its channel functions 
possess the symmetry property 
\begin{equation}
\varphi_{n,\alpha}(x) = \text{S}_{\alpha} \, (-1)^{n} \varphi_{n,\alpha}(-x) 
\label{Eq:Floqet_channel_symmetry}
\end{equation}
according to the Floquet-state category, 
that is: 
(i) for a Floquet eigenstate of S$_{\alpha}$=($+$), 
$\varphi_{n,\alpha}(x)$=$\mp\varphi_{n,\alpha}(-x)$ for odd and even $n$, 
respectively; 
whereas
(ii) for a Floquet eigenstate of S$_{\alpha}$=($-$),  
$\varphi_{n,\alpha}(x)$=$\pm\varphi_{n,\alpha}(-x)$ for odd and even $n$, 
respectively.


The temporal amplitude of the emitted (HH) field 
is proportional to 
the time-dependent electron acceleration~\cite{Nimrod_Weinhold_PRL_1997,Nimrod_Lein_JPhysChemA_2003,Nimrod_Aver_PRA_2008}. 
%
Hence, when the system populates a single given Floquet eigenstate $\alpha$, 
it is obtained from 
\begin{equation}
a_{\alpha}^{\text{HH}}(t) = \frac{\partial^2}{\partial t^2}  \left( \chi_{\alpha}(x,t) \right|  x  \left| \chi_{\alpha}(x,t) \right) \; .
\end{equation}
Following Fourier transform, 
the resulting relative amplitude of the high-harmonic field 
generated at frequency $\Omega = M \omega_{0}$,
with $M$ being an integer (for simplicity),
%
%
is then given by~\cite{Nimrod_Weinhold_PRL_1997,Nimrod_Aver_PRA_2008}
\begin{equation}
A_{\alpha}^{\text{HH}}(M\omega_{0}) = - (M\omega_{0})^{2} 
\sum_{n=-\infty}^{\infty}\left(\varphi_{n+M,\alpha}(x)\right| x \left|
\varphi_{n,\alpha}(x)\right)_{x} ,
\label{Eq:A_alpha_HHS}
\end{equation}
%
%
%
i.e., it results from a summation over all the dipole matrix elements between pairs of Floquet  channel functions 
of eigenstate $\alpha$
whose orders differ by the harmonic order $M$.
The corresponding high-harmonic spectral intensity is then 
\begin{equation}
I_{\alpha}^{\text{HH}}(M\omega_{0}) = \left| A_{\alpha}^{\text{HH}}(M\omega_{0}) \right|^{2} \; .
\label{Eq:I_alpha_HHS}
\end{equation}
%
%
%
Following Eqs.~(\ref{Eq:Floqet_state_symmetry})$-$(\ref{Eq:I_alpha_HHS}), 
symmetry considerations directly imply
that 
$A_{\alpha}^{\text{HH}}(M\omega_{0})$ and
$I_{\alpha}^{\text{HH}}(M\omega_{0})$ 
are generally nonzero for odd $M$ and zero for even $M$,
i.e., only odd high harmonics are generated when a 
single Floquet eigenstate is populated~\cite{Nimrod_Weinhold_PRL_1997,Nimrod_Lein_JPhysChemA_2003,Nimrod_Aver_PRA_2008}.
%
%
%
It applies to both the so-called 
plateau and cutoff regions of the HH spectrum. 

\begin{figure*} 
\includegraphics[width=0.75\columnwidth,center]{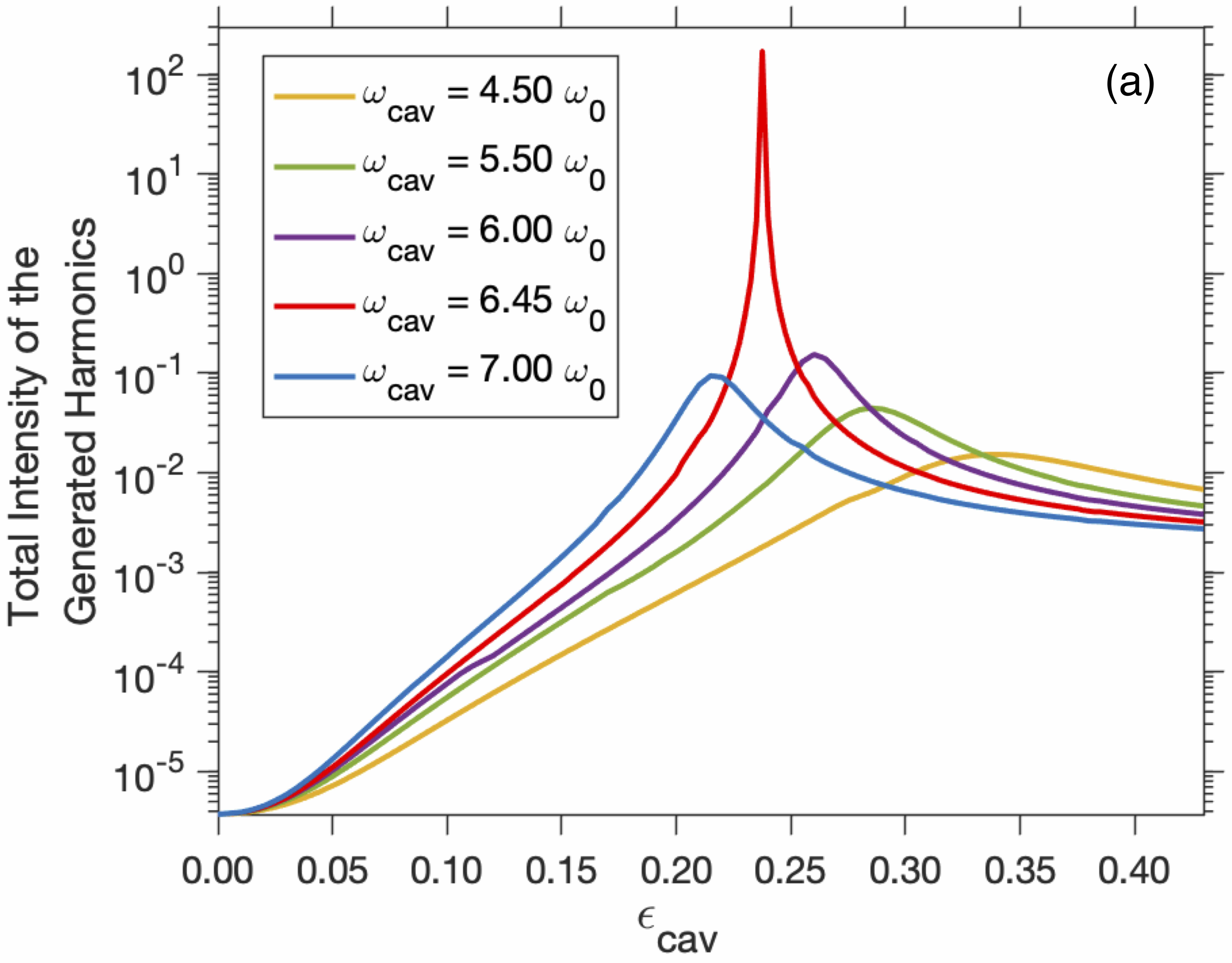}
\includegraphics[width=0.5\columnwidth]{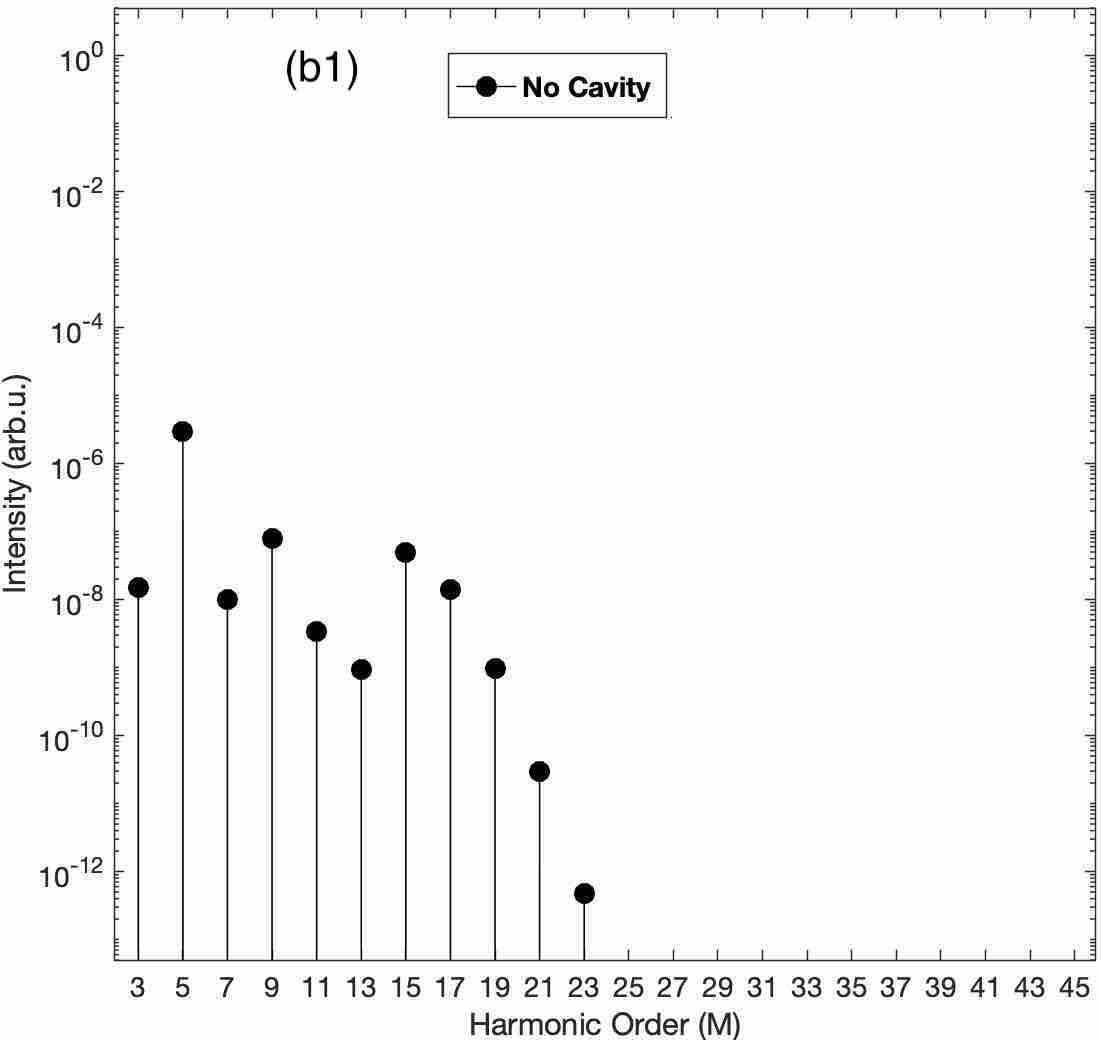}
\includegraphics[width=0.5\columnwidth]{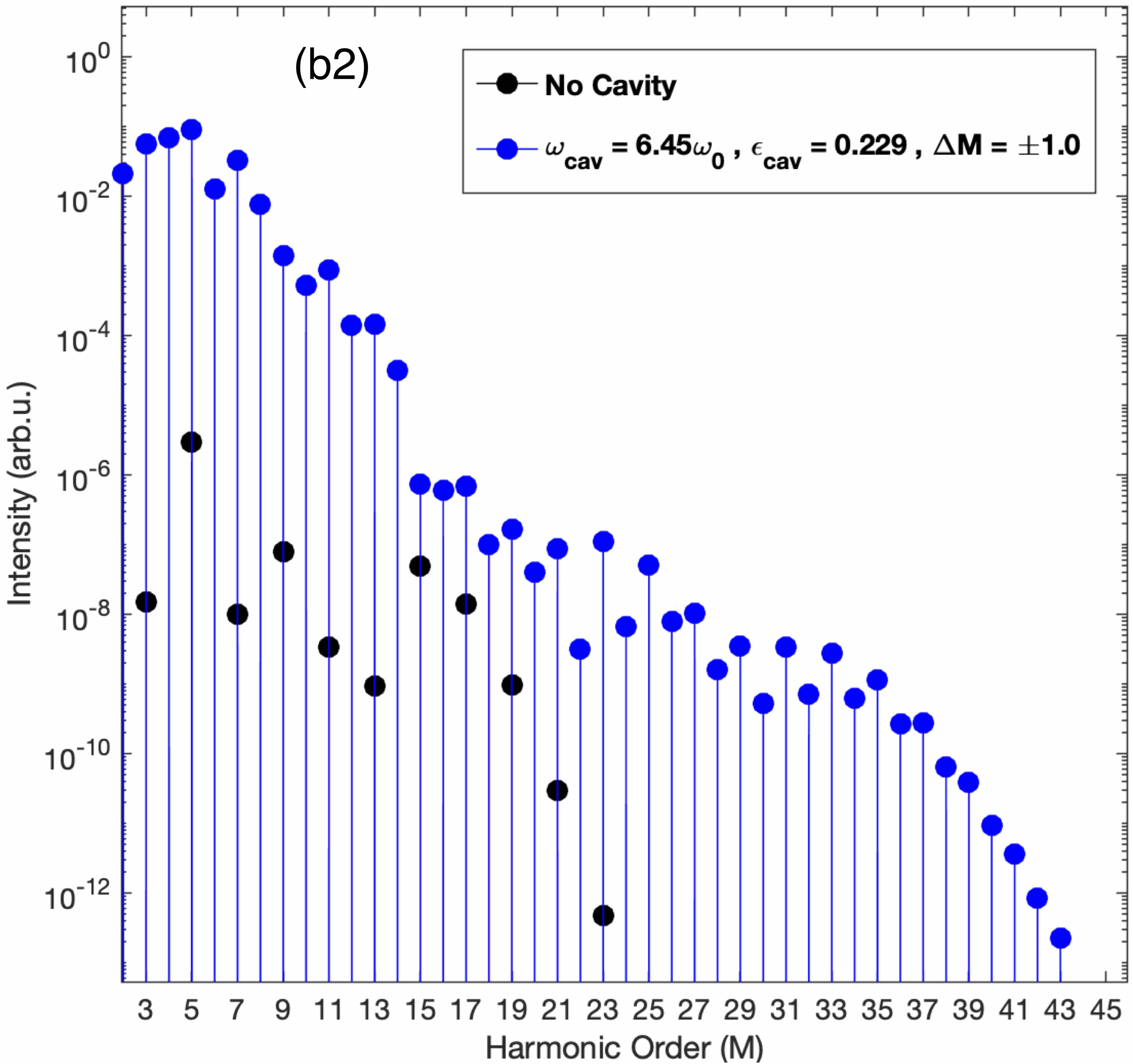}
\includegraphics[width=0.5\columnwidth]{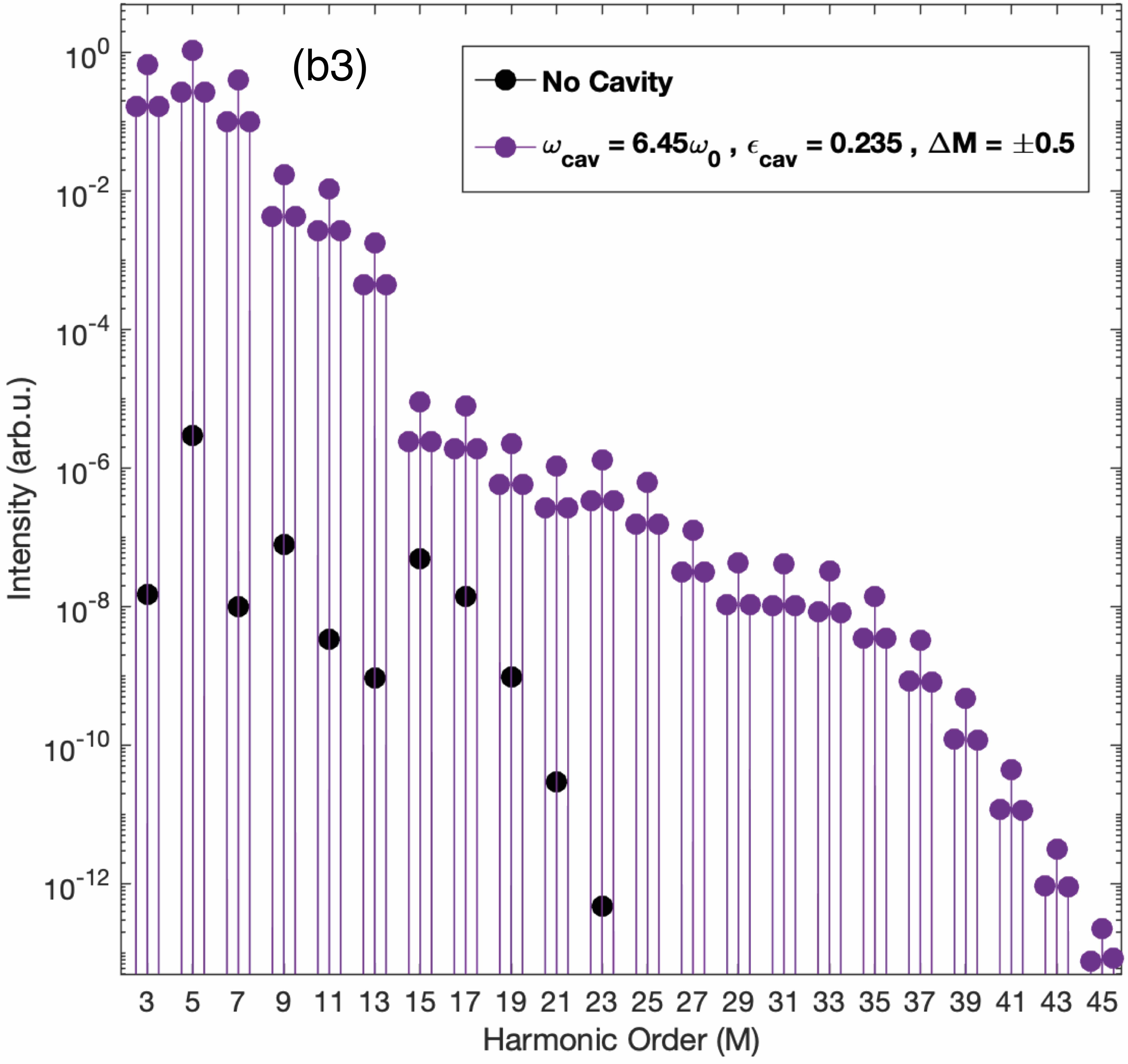}
\includegraphics[width=0.5\columnwidth]{fig-spacing-low-res.jpg}
\includegraphics[width=0.5\columnwidth]{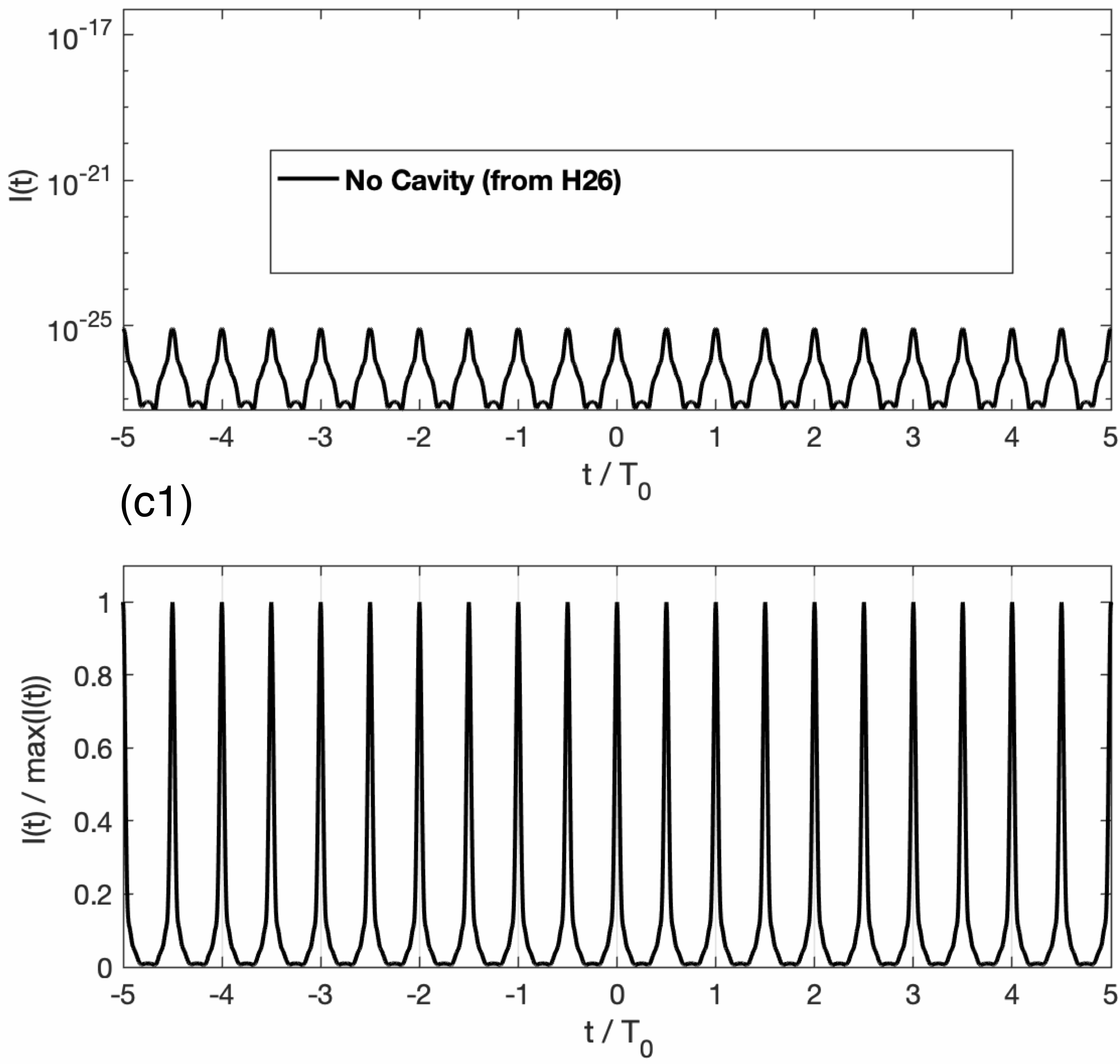}
\includegraphics[width=0.5\columnwidth]{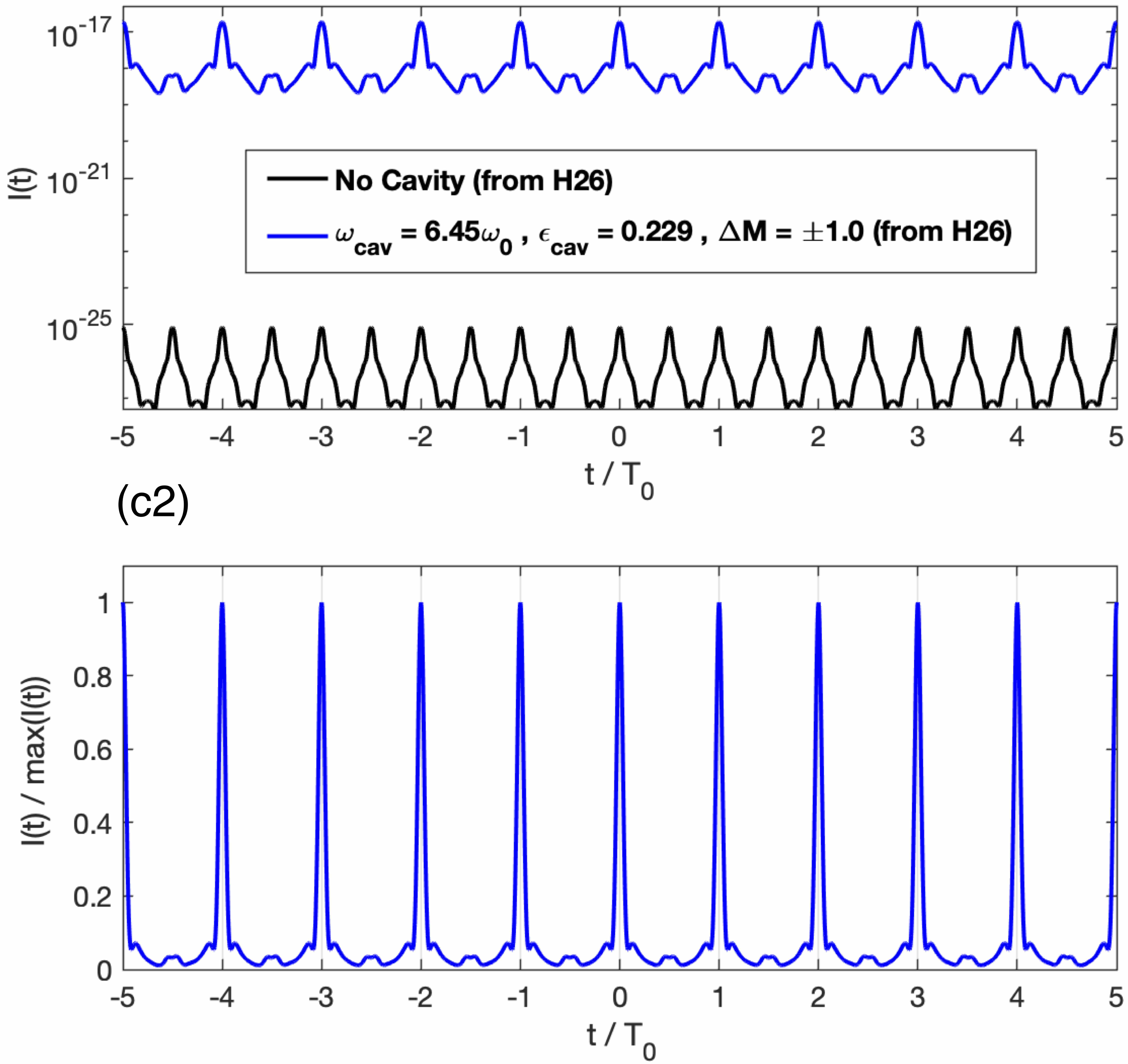}
\includegraphics[width=0.5\columnwidth]{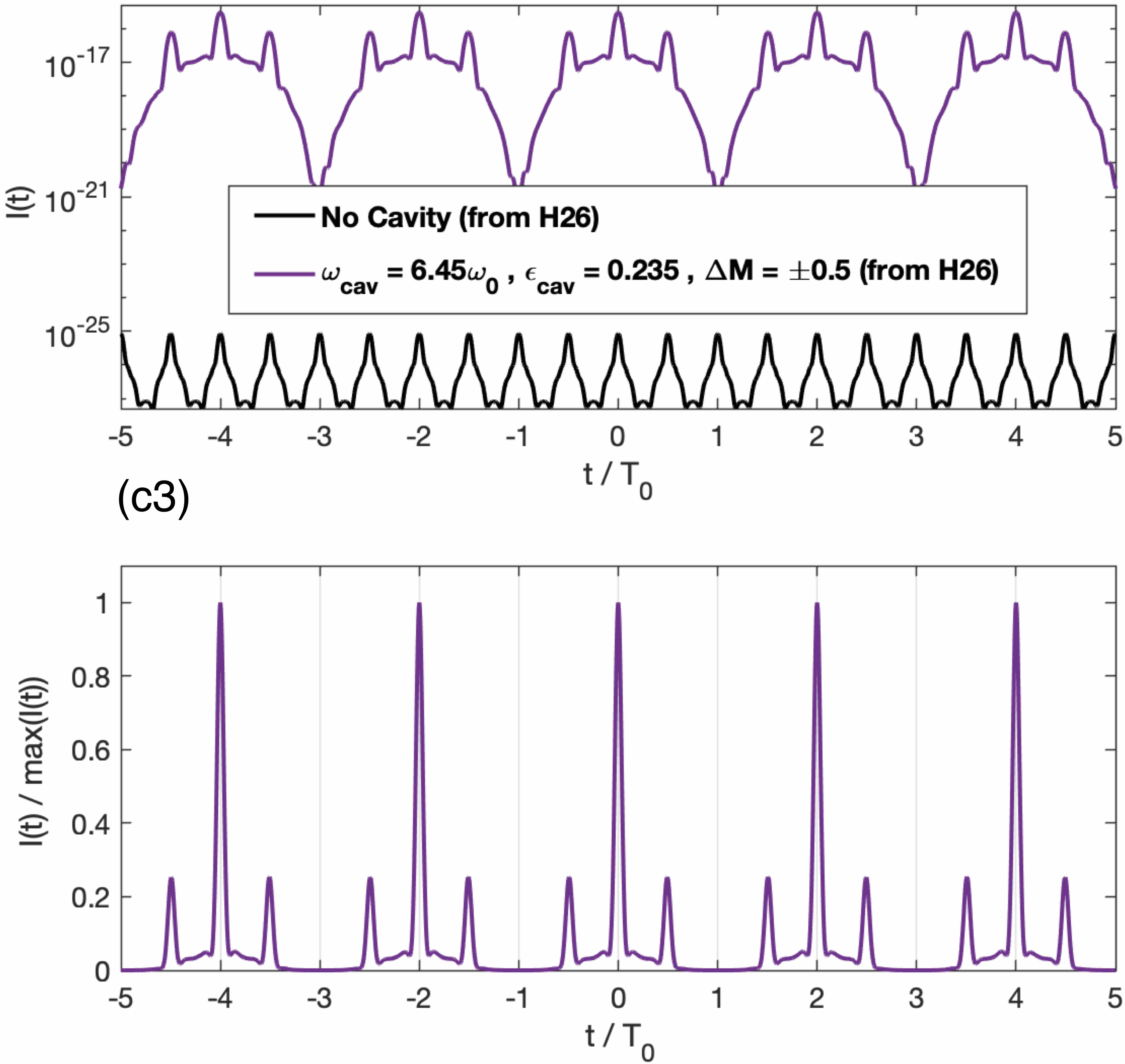}
\includegraphics[width=0.5\columnwidth]{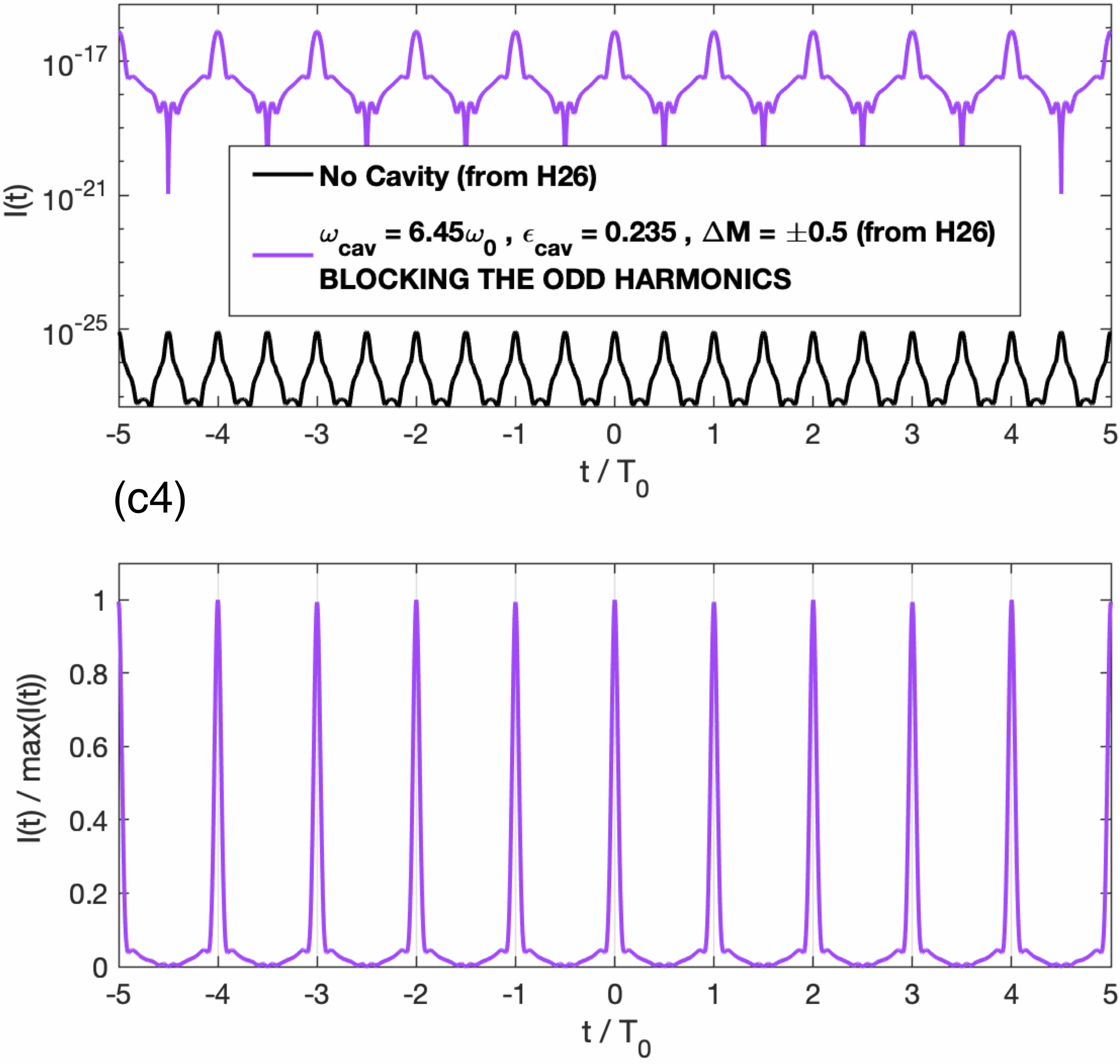}
\includegraphics[width=0.5\columnwidth]{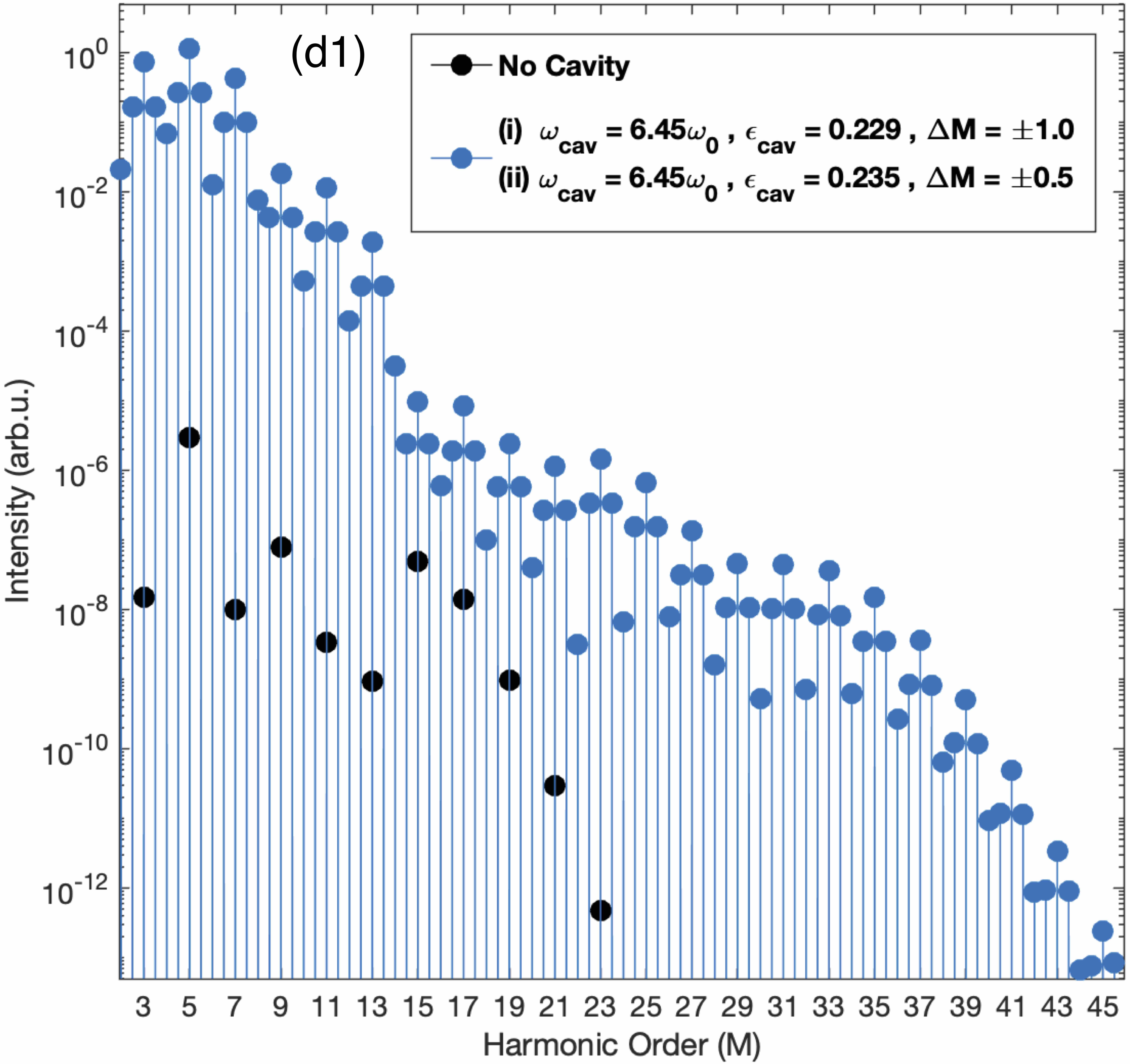}
\includegraphics[width=0.5\columnwidth]{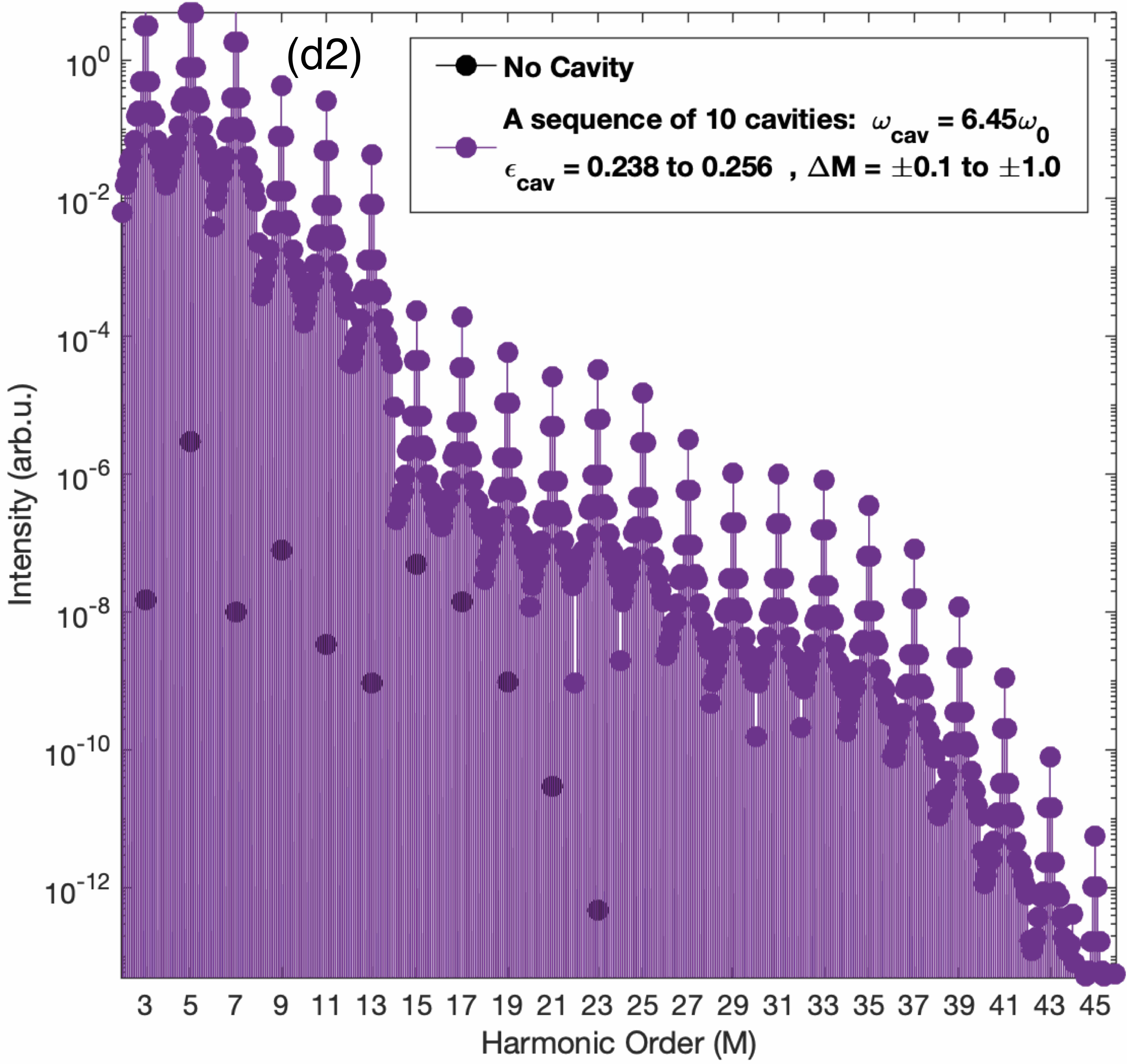}
\caption{
Examples of the theoretical results numerically calculated 
for the cavity-controlled HHG by single irradiated Xe atoms in one or more cavities.
Each of the cavities is initially seeded with a single photon.
Each of the Xe atoms is initially in its field-free ground state, and is irradiated by a strong linearly-polarized cw field having a frequency of $\omega_{0}$=0.057  
(a wavelength of 800~nm) 
and an amplitude of $\varepsilon_{0}$=0.04 
(an intensity of 5$\times$10$^{13}$~W/cm$^{2}$).
\\
(A) For the HH yield generated by a single irradiated Xe atom in a single cavity:
Panel~(a) shows the total HH intensity 
as a function of $\varepsilon_{cav}$ 
for several values of $\omega_{cav}$. 
%
\\
(B) For the HH spectrum generated by a single irradiated Xe atom in a single cavity:
Panel~(b1) shows the generated HH spectrum for the case of no cavity,
and 
panels~(b2) and (b3) show the generated HH spectrum for, respectively, 
the cases of:  
(i) $\omega_{cav}$=6.45$\omega_{0}$ and $\varepsilon_{cav}$ = 0.229, 
corresponding to a side-harmonic shift of $\Delta M$=$\pm$1.0,
and
(ii) $\omega_{cav}$=6.45$\omega_{0}$ and $\varepsilon_{cav}$ = 0.235, 
corresponding to a side-harmonic shift of $\Delta M$=$\pm$0.5.
\\
(C) For the attosecond pulse sequence that can be generated in the different HHG cases above:
Panel~(c1) shows the TL temporal attosecond pulse sequence generated 
from the cutoff region (beyond the 26th harmonic; H26) 
of the no-cavity HH spectrum presented in 
panel~(b1),
after setting all the HH spectral phase to zero.
%
Panels~(c2) and (c3) show the TL attosecond pulse sequence generated from the cutoff region (beyond H26) of the cavity-induced HH spectrum presented in 
panels~(b2) and (b3), respectively.
%
%
Panel~(c4) shows the TL attosecond pulse sequence 
generated from the cutoff region (beyond H26) of the cavity-induced HH spectrum presented in panel~(b3), 
after the odd harmonics are blocked. 
%
\\
(D) Panel~(d1) presents results for the HH spectrum generated by a sequence of two cavities
with a single irradiated Xe atom in each.
Specifically here, these two cavities are the individual cavities of 
panels~(b2) and (c2),
one corresponding to $\Delta M$=$\pm$1.0 while the other to $\Delta M$=$\pm$0.5.
%
Panel~(d2) displays results for the HH spectrum generated by a sequence of ten cavities
with a single irradiated Xe atom in each.
All the ten cavities have a frequency of $\omega_{cav}$=6.45$\omega_{0}$,
while their $\varepsilon_{cav}$ spans the range of 0.238 to 0.256,
corresponding to shift magnitude $\left| \Delta M \right|$ of 
0.1 to 1.0 in a step of 0.1. 
}    
\label{fig:results_HHG_single_atoms}
\end{figure*}

Next, we consider a cavity-free system of a 
Xe atom that initially populates its field-free ground (electronic) state and is then irradiated by a strong cw field 
having an amplitude of up to about $\varepsilon_{0}$=0.4 (atomic units).
%
%
As previously shown~\cite{Nimrod_Weinhold_PRL_1997}, 
this system 
predominantly populates the Floquet eigenstate $\alpha$=$FLg$ 
that its eigenfunction 
$\left| FLg \right) = \left| \Phi_{FLg}(x,t) \right)$
has the largest spatial overlap with the 
the eigenfunction of the field-free ground state of Xe 
among all the system's Floquet eigenfunctions. 
%
%
%
%
This largest overlap is of value close to 1.
Hence, as presented above, only odd high harmonics are generated during such cw irradiation of Xe.
%
%
This is the non-Hermitian Floquet quantum-mechanical 
explanation for this well-known 
phenomenon of odd-only HHG 
by ground-state atoms irradiated by a cw field~\cite{HHG_paper1_Ferray_1988,HHG_paper2_LHuillier_1991,HHG_paper3_Kulander_1992,Nimrod_Weinhold_PRL_1997,Nimrod_Lein_JPhysChemA_2003,Nimrod_Aver_PRA_2008}.
%
%
%
Fig.~\ref{fig:results_HHG_single_atoms}(b1) 
shows the corresponding cavity-free  
theoretical results 
of the odd-only HH spectrum 
calculated for 
the model ground-state Xe atom irradiated by a cw field of 
amplitude $\varepsilon$=0.04, 
corresponding to an intensity of 5$\times$10$^{13}$~W/cm$^{2}$, 
and of frequency $\omega_{0}$=0.057, 
corresponding to a wavelength of 800~nm.
%
Throughout this paper, 
the units of quantities written as unitless are atomic units.

\begin{figure} 

%
\includegraphics[width=0.90\columnwidth,center]{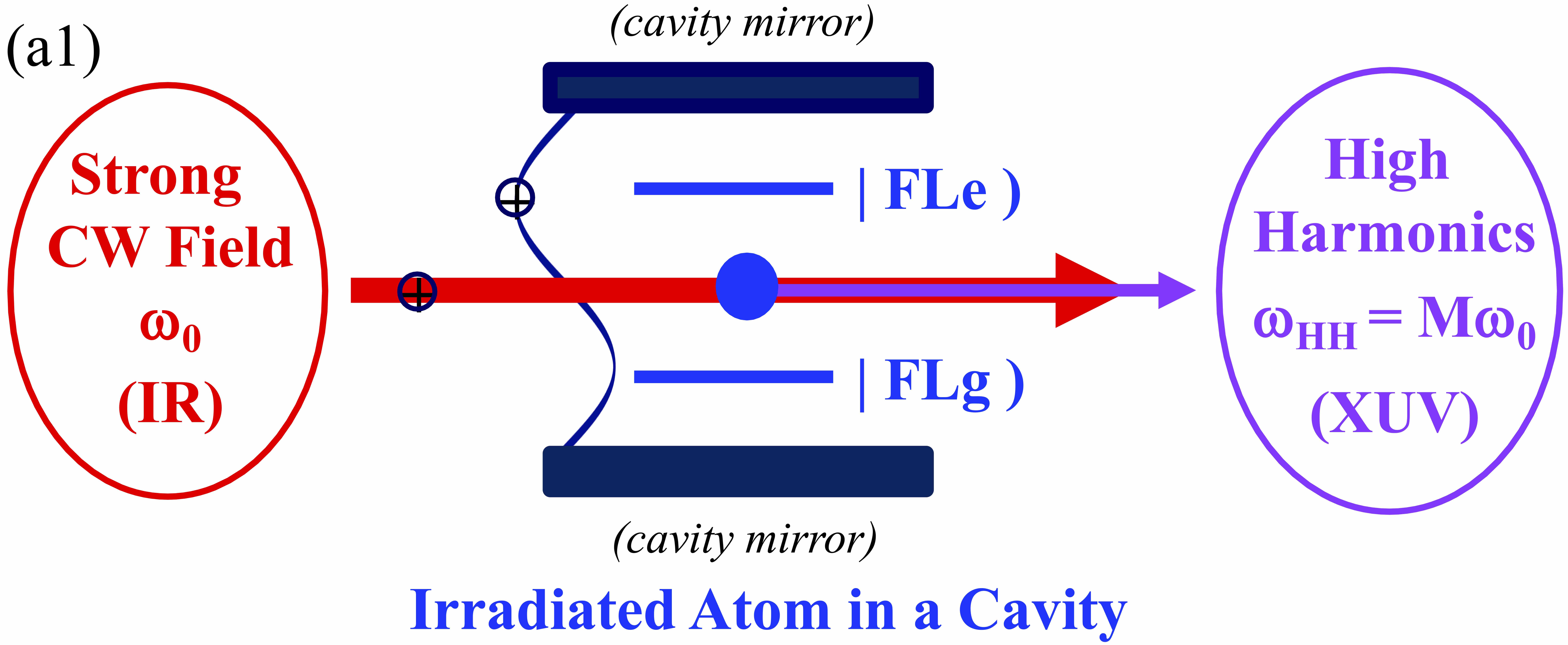}

\includegraphics[width=0.01\columnwidth,center]{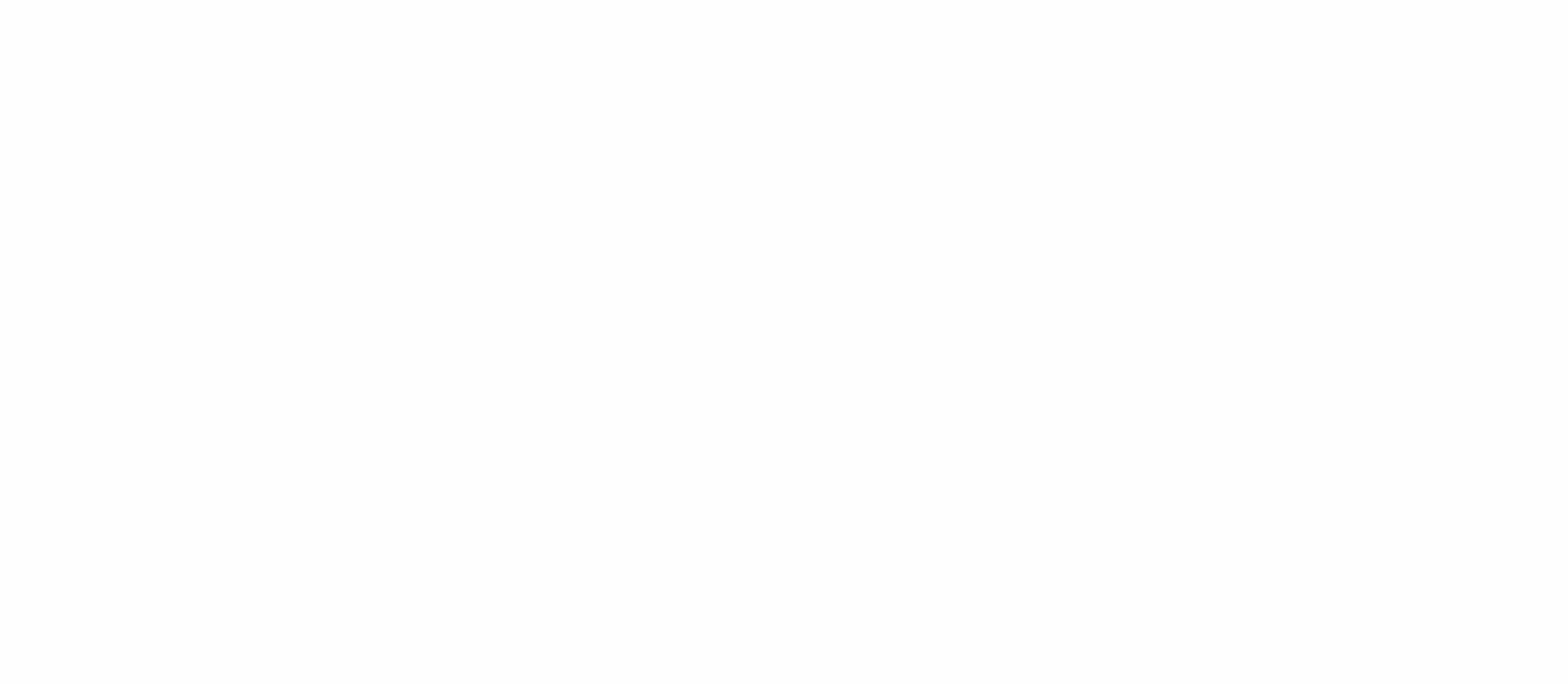}


\includegraphics[width=0.95\columnwidth,center]{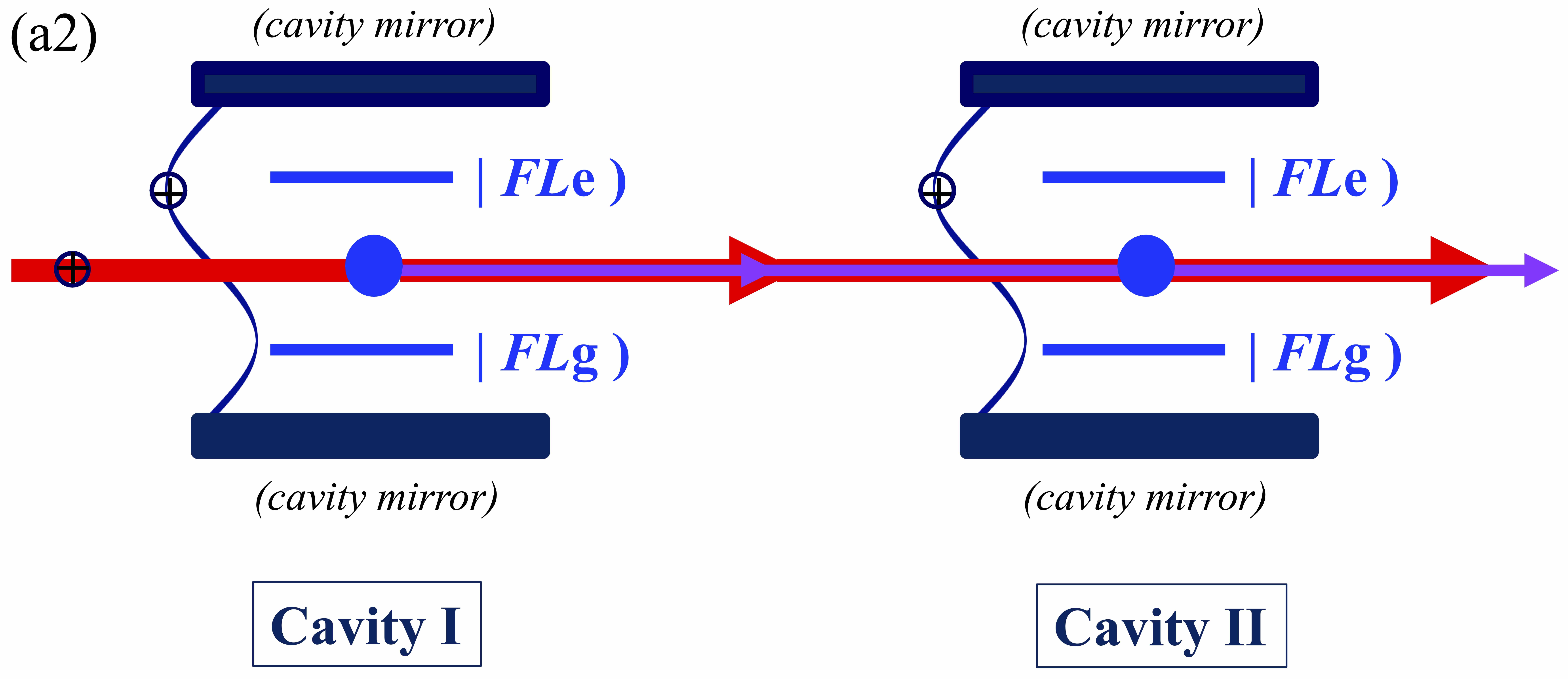}

\includegraphics[width=0.12\columnwidth,center]{fig-spacing-low-res.jpg}



\caption{Schematic of the general scheme for high harmonic generation by individual irradiated atoms in cavities. See text for details.}

\label{fig:general_scheme}

\end{figure}

%
%
Following the well-established framework described above, we move now 
to the innovative part of the study, presenting the cavity-controlled HHG.
Here, the previously described system$-$an irradiated Xe atom initially in its field-free ground state and then in the 
cavity-free Floquet eigenstate $FLg$ (see above)$-$is placed within a single-mode quantum cavity
that is initially seeded with a single photon.
%
%
%
%
%
%
%
Fig.~\ref{fig:general_scheme}(a1) schematically shows the corresponding scheme.
%
%
As considered earlier, 
a strong classical cw field   
with amplitude $\varepsilon_{0}$ and frequency $\omega_{0}$,
linearly polarized along the $x$ direction 
and propagating along the $z$ direction,  
irradiates the model Xe atom having its atomic coordinate $x$ along the $x$ direction.  
%
%
In addition, the newly introduced cavity 
of frequency $\omega_{cav}$ 
has its axis along the $y$ direction, and 
its single-mode field considered here
is also linearly polarized along the $x$ direction.
This configuration ensures, for clarity and simplicity, that the photonic modes of the cavity and of the strong cw field are different.
Additionally, for simplicity, to avoid the possibility of any emitted 
high-harmonic radiation being trapped in the cavity, $\omega_{cav}$ is assumed to be a non-integer multiple of $\omega_{0}$.


The Hamiltonian $\hat{H}_{cav}$ describing 
this whole system of the irradiated Xe atom in a cavity is 
then given by the cavity-free $\hat{H}_{}$ [Eq.(\ref{eq:H})] 
with the addition of time-independent 
cavity-dependent terms, 
\begin{align}
    \hat{H}_{cav} = \, 
    & \hat{H}_{}(x,t) \; + \nonumber \\
    & \hbar\omega_{cav} \, \hat{a}^{\dagger}_{cav}\hat{a}_{cav} + 
    \hat{d} \, \varepsilon_{cav} \,(\hat{a}^{\dagger}_{cav}+\hat{a}_{cav}) ,  
\label{eq:H_cav}     
\end{align}
where $\hat{a}^{\dagger}_{cav}$ and $\hat{a}_{cav}$ are the 
creation and annihilation operators of the cavity mode, respectively.
The quantity 
$\varepsilon_{cav}$$\equiv$
$ \sqrt{\left(\hbar\omega_{cav}\right) / 
\left(2 \Bar\epsilon_{0} V\right)}$ 
is the cavity-atom coupling strength parameter,
with $V$ being the cavity (quantization) volume 
and $\Bar\epsilon_{0}$ being the vacuum permittivity.
The 
dipole self-energy term 
$\left[ \varepsilon_{cav}^{2} {\hat{d}_{\theta}}^2 / (\hbar\omega_{cav}) \right]$  
has been omitted from $\hat{H}_{cav}$.
%
%
%
%
%
The corresponding non-Hermitian Floquet Hamiltonian $\hat{H}_{cav,f}$ 
is then given by the cavity-free $\hat{H}_{f}$ [Eq.(\ref{eq:H_f})] with the addition of the corresponding cavity-dependent terms,  
\begin{align}
    \hat{H}_{cav,f} = \, 
     & \hat{H}_{f}(x_{\theta},t) \; + \nonumber \\ 
     & \hbar\omega_{cav} \, \hat{a}^{\dagger}_{cav}\hat{a}_{cav} + \hat{d}_{\theta} \, \varepsilon_{cav} \,(\hat{a}^{\dagger}_{cav}+\hat{a}_{cav}) .  
\label{eq:H_f_cav}     
\end{align}
The eigenstates of $\hat{H}_{cav,f}$ are 
Floquet eigenstates that simultaneously describe the irradiated Xe atom and the cavity field; 
we therefore term them polaritonic Floquet eigenstates (Floquet polaritons).
The initial (polaritonic) state of the system$-$here, 
an irradiated Xe atom in 
$FLg$, 
and a cavity with a single photon$-$determines their amplitudes. 
This, in turn, dictates the time-dependent wavefunction of the system and 
the emitted HH field.
%
Overall, the generated HH spectrum 
is determined and controlled 
by the complete set of properties 
of the different system components$-$i.e., the atom, the strong cw field,   
and the cavity$-$as well as by the initial state of the system.

In analogy to the Jaynes-Cummings model~\cite{JaynesCummingsModel}, 
for studying the essence of 
cavity-controlled HHG, 
we consider here a model that includes 
only a pair of cavity-free Floquet eigenstates of the irradiated Xe atom.
%
%
Specifically, 
they are chosen as the cavity-free Floquet eigenstates 
$\alpha$=$FLg$ and $\alpha$=$FLe$ that 
%
%
have the largest spatial overlap, 
among all the cavity-free Floquet eigenstates,
with the field-free ground and first-excited states of Xe, respectively.
The real parts of their quasienergies, 
$\epsilon_{_{FLg}}$ and $\epsilon_{_{FLe}}$,
satisfy $E_{FLg} < E_{FLe}$.
%
%
As described before, 
the initial (polaritonic) wavefunction of the system is 
\begin{equation}
\left| \Psi_{cav}(t=0) \right) = \left| FLg, 1 \right) = 
\left| \Phi_{FLg}(x,t) \right) \, \left|1\right>_{cav},
\end{equation}
where $\left|n_{ph}\right>_{cav}$ is the photonic eigenstate (QED number state) of the cavity mode with $n_{ph}$ photons and no atom in the cavity. 
%
The present study is conducted under the Rotating-Wave Approximation.
Therefore, following this $\Psi_{cav}(t=0)$,  
the system and its dynamics are fully described within the 
"single-excitation" polaritonic subspace 
spanned by the two basis functions
\begin{eqnarray}
\left| FLg, 1 \right) & = & \left| \Phi_{FLg}(x,t) \right) \, \left|1\right>_{cav} 
\nonumber \\
\left| FLe, 0 \right) & = & \left| \Phi_{FLe}(x,t) \right) \, \left|0\right>_{cav} \, ,
\nonumber
\end{eqnarray}
which correspond to the irradiated Xe atom in either $FLg$ or $FLe$, with the cavity mode containing one or zero photons, respectively.

Hence, $\hat{H}_{cav,f}$ is given by the following 2$\times$2 matrix representation:
\begin{align*}
& \left[\hat{H}_{cav,f}\right]_{11} =   
\left( FLg,1 \left| \hat{H}_{cav,f} \right| FLg,1 \right) = 
\nonumber \\
& = \epsilon_{_{FLg}} + \hbar \omega_{cav}  = \left( E_{_{FLg}} - i\Gamma_{_{FLg}}/2 \right) + \hbar \omega_{cav} \, , 
\nonumber \\ \nonumber \\
& \left[\hat{H}_{cav,f}\right]_{22} = 
\left( FLe,0 \left| \hat{H}_{cav,f} \right| FLe,0 \right) = &
\nonumber \\
& = \epsilon_{_{FLe}} = \left( E_{_{FLe}} - i\Gamma_{_{FLe}}/2 \right) \, ,
\nonumber \\ \nonumber \\
& \left[\hat{H}_{cav,f}\right]_{12} = 
\left( FLe,0 \left| \hat{H}_{cav,f} \right| FLg,1 \right) = &
\nonumber \\
& = \varepsilon_{cav} \, d_{_{FLe,FLg}}
= \varepsilon_{cav} \sum_{n=-\infty}^{\infty}\left(\varphi_{n,_{FLe}}(x) \right| x \left| \varphi_{n,_{FLg}}(x)\right)_{x} \, ,
\nonumber \\ \nonumber \\
& \left[\hat{H}_{cav,f}\right]_{21} =  
\left( FLg,1 \left| \hat{H}_{cav,f} \right| FLe,0 \right) = &
\nonumber \\
& = \varepsilon_{cav} \, d_{_{FLg,FLe}}
= \varepsilon_{cav} \sum_{n=-\infty}^{\infty}\left(\varphi_{n,_{FLg}}(x) \right| x \left| \varphi_{n,_{FLe}}(x)\right)_{x} \, .
%
\end{align*}
Following 
Eqs.~(\ref{Eq:Floqet_state_symmetry})$-$(\ref{Eq:Floqet_channel_symmetry}), 
the non-diagonal terms,
$\left[\hat{H}_{cav,f}\right]_{12}$ and
$\left[\hat{H}_{cav,f}\right]_{21}$,
are non-zero only when $FLg$ and $FLe$ differ in their dynamical symmetry; 
as analyzed below, only in such a case does the cavity have any influence on the HHG.
The polaritonic Floquet eigenstates of the system within the single-excitation subspace, i.e., the upper and lower Floquet polaritons,
are then
given as the eigenstates of $\hat{H}_{cav,f}$,
\begin{equation}
{\Large{\hat{H}}}_{cav,f} \, 
\left| {\Large{\Phi}}_{cav,\pm}^{(\omega_{cav},\varepsilon_{cav})} \right) =  {\Large{\epsilon}}_{cav,\pm}^{(\omega_{cav},\varepsilon_{cav})} \, 
\left| {\Large{\Phi}}_{cav,\pm}^{(\omega_{cav},\varepsilon_{cav})} \right) \; ,
\end{equation}
with their 
(complex) eigenenergies being   
\begin{equation}
{\Large{\epsilon}}_{cav,\pm}^{\;(\omega_{cav},\varepsilon_{cav})} =  \frac{1}{2} \left( \epsilon_{_{FLg}} + \hbar \omega_{cav} +  \epsilon_{_{FLe}} \right) \, \pm \, \frac{1}{2} \hbar \, \Omega \; ,
\end{equation}
and their 
eigenfunctions being
\begin{align}
\left| {\Large{\Phi}}_{cav,\pm}^{(\omega_{cav},\varepsilon_{cav})} \right) & = 
\sqrt{\frac{\Omega \pm \delta}{2 \Omega} } \left| FLg,1 \right) \pm \sqrt{\frac{\Omega \mp \delta}{2 \Omega} } \left| FLe,0 \right) \; ,
\label{eq:Phi_cav}
\end{align}
where 
\begin{align}
& \hbar \, \Omega = \sqrt{ (\hbar \, \delta)^{2} + (\hbar \, \Omega_{0})^{2} } \; ,
\nonumber \\ \nonumber \\ 
& \hbar \, \delta = \epsilon_{_{FLg}} + \hbar \omega_{cav} - \epsilon_{_{FLe}} \; ,
\nonumber \\ \nonumber \\ 
& \hbar \, \Omega_{0} = 2 \, \varepsilon_{cav} \, d_{_{FLg,FLe}} = 
\nonumber \\
& \hspace{0.7cm} = 2 \, \varepsilon_{cav} \sum_{n=-\infty}^{\infty}\left(\varphi_{n,_{FLe}}(x) \right| x \left| \varphi_{n,_{FLg}}(x)\right)_{x} \; ,
\end{align}
with $\Omega_{0}$ being the system's (complex) Rabi frequency.

With these upper and lower Floquet polaritons,  
the time-dependent wavefunction of the system
evolves from its initial state
$\Psi_{cav}(t$=0)=$\left| FLg, 1 \right)$
as their coherent superposition:
\begin{align}
& \big| \Large{\Psi}_{cav}(t) \big) = \;\;
\nonumber \\ 
& \hspace{0.8cm} = \Large{a}_{cav,+}  \; 
\exp\left(-i {\Large{\epsilon}}_{cav,+}^{\;(\omega_{cav},\varepsilon_{cav})}  t /\hbar\right) \; 
\left| {\Large{\Phi}}_{cav,+}^{(\omega_{cav},\varepsilon_{cav})} \right) \; + 
\nonumber \\ 
& \hspace{1.3cm}\Large{a}_{cav,-} \; 
\exp\left(-i {\Large{\epsilon}}_{cav,-}^{\;(\omega_{cav},\varepsilon_{cav})}  t /\hbar \right) \; \left| {\Large{\Phi}}_{cav,-}^{(\omega_{cav},\varepsilon_{cav})} \right) \; ,
\nonumber \\ \nonumber \\ 
& \Large{a}_{cav,\pm} = \sqrt{\frac{\Omega \pm \delta}{2 \Omega} } \; .
\label{eq:Psi_irAtom_cav}
\end{align}
The generated HH field is then given by 
\begin{equation}
a_{cav}^{\text{HH}}(t) = \frac{\partial^2}{\partial t^2}  
\left( \Large{\Psi}_{cav}(t) \right|  x  \left| \Large{\Psi}_{cav}(t) \right) \; .
\label{eq:accelataion_cav}
\end{equation}
%
%

Using Eq.~(\ref{eq:Psi_irAtom_cav}), 
the right-hand side of Eq.~(\ref{eq:accelataion_cav}) contains terms 
that involve either 
$\Phi_{cav,+}^{(\omega_{cav},\varepsilon_{cav})}$ or $\Phi_{cav,-}^{(\omega_{cav},\varepsilon_{cav})}$ alone 
(see Eq.~(\ref{eq:Phi_cav})), 
and others that involve both.
Following Fourier transform, these two groups of terms generally give rise to different parts of the resulting cavity-controlled HH spectrum: the former yields odd harmonics, which also exist in the no-cavity case, whereas the latter generates side harmonics around these odd harmonics, 
which do not appear without the cavity.
The respective HH field amplitudes 
at frequencies $\Omega = M \omega_{0}$ and $\Omega = (M \pm \Delta M) \omega_{0}$, 
with $M$ being an integer, are obtained to be 
\begin{align}
& A_{cav}^{\text{HH}}\Big( M\omega_{0} \Big) = 
\nonumber \\
& = \left(\frac{\Omega^2 + \delta^2}{2 \Omega^2}\right)  
A_{FLg}^{\text{HH}}(M\omega_{0}) +
\left(\frac{\Omega^2 - \delta^2}{2 \Omega^2}\right)   
A_{FLe}^{\text{HH}}(M\omega_{0}) \; , 
\label{eq:A_cav_odd_harm}
\end{align}
and 
\begin{align}
& A_{cav}^{\text{HH}}\Big( (M \pm \Delta M) \omega_{0} \Big) = 
\nonumber \\ 
& = \kappa 
\left(\frac{\Omega^2 - \delta^2}{4 \Omega^2}\right)
\Big( A_{FLg}^{\text{HH}}(M\omega_{0}) - A_{FLe}^{\text{HH}}(M\omega_{0}) \Big)
\; , \nonumber \\ \nonumber \\ 
& \Delta M = 
\left( \Large{\epsilon}_{cav,+}^{\;(\omega_{cav},\varepsilon_{cav})} - 
             \Large{\epsilon}_{cav,-}^{\;(\omega_{cav},\varepsilon_{cav})} \right) / 
      \left( {\hbar \, \omega_{0}}  \right) \;  , 
\nonumber \\ \nonumber \\ 
& \kappa = \left(\frac{M \pm \Delta M}{M}\right)^{2} \; , \nonumber \\
\label{eq:A_cav_odd_side_harm}
\end{align}
where
$A_{FLg}^{\text{HH}}(M\omega_{0})$ and $A_{FLe}^{\text{HH}}(M\omega_{0})$
are the cavity-free high-harmonic field amplitudes
generated at frequency $\Omega = M \omega_{0}$ 
when the irradiated Xe atom 
populates solely 
$FLg$ or $FLe$, respectively 
(see Eq.~(\ref{Eq:A_alpha_HHS})).
%
The HH amplitude generated in the cavity at frequency $\Omega$ 
is then generally given by 
\begin{align}
& A_{cav}^{\text{HH}}\Big( \Omega \Big) = \
\delta_{\Omega,M\omega_{0}} \; A_{cav}^{\text{HH}}\Big( M\omega_{0} \Big) \ + 
\nonumber \\
& \hspace{2.0cm} \delta_{\Omega,(M \pm \Delta M) \omega_{0}} \;
A_{cav}^{\text{HH}}\Big( (M \pm \Delta M) \omega_{0} \Big) \; , 
\end{align}
with the corresponding HH spectral intensity being 
\begin{equation}
I_{cav}^{\text{HH}}\Big( \Omega \Big) =
\left| A_{cav}^{\text{HH}}\Big( \Omega \Big) \right|^{2} \; .
\end{equation}
As seen, for a given atomic system and strong cw field, 
the cavity parameters $\omega_{cav}$ and $\varepsilon_{cav}$ 
determine the shift parameter $\Delta M$, 
which sets the displacement of the side harmonics 
at frequencies $(M \pm \Delta M) \omega_{0}$ 
relative to the odd harmonic at frequency $M\omega_{0}$.

Following the above, the cavity control over the generated HH spectrum can be further enhanced by extending the basic scheme of a single irradiated Xe atom in a single cavity to an arrangement of several ($N_{cav}$) cavities placed in a row, each containing one irradiated Xe atom. 
Each Xe atom is initially in its field-free ground electronic state, and each cavity is seeded with a single photon. 
The strong cw field passes through the sequence of cavities, accompanied by the HH fields generated in all preceding cavities, which coherently add together.  
As an example, the case of two cavities in a row is schematically illustrated in Fig.~\ref{fig:general_scheme}(a2).
With each of the $N_{cav}$ cavities 
generally having a different pair of values for $\omega_{cav}$ and $\varepsilon_{cav}$
that corresponds to a different shift parameter $\Delta M$
(i.e., cavity $i$ produces side harmonics with a shift of $\Delta M_{i}$),
the overall HH spectrum contains now a series of side harmonics around each odd harmonic 
at the shifts $\Delta M_{1}, \Delta M_{2}, \dots, \Delta M_{N_{cav}}$.

Figure~\ref{fig:results_HHG_single_atoms} shows 
several illustrative examples of the corresponding theoretical results 
numerically calculated 
for the cavity-controlled HHG by single irradiated Xe atoms in one or more cavities.
Each of the cavities is initially seeded with a single photon.
Each of the Xe atoms is initially in its field-free ground state, and is irradiated by a strong linearly-polarized cw field having a frequency of $\omega_{0}$=0.057  
(corresponding to a wavelength of 800~nm) 
and an amplitude of $\varepsilon_{0}$=0.04 
(corresponding to an intensity of 5$\times$10$^{13}$~W/cm$^{2}$).
For the HH yield generated by a single irradiated Xe atom in a single cavity,
Fig.~\ref{fig:results_HHG_single_atoms}(a) shows the total HH intensity 
$I^{HH}_{tot}$ 
(i.e., after summation over all frequencies) 
as a function of $\varepsilon_{cav}$ 
for several values of $\omega_{cav}$. 
Note the logarithmic scale of the y-axis.
As seen, there is a very high sensitivity of $I^{HH}_{tot}$ to the cavity parameters,  
with orders-of-magnitude changes among the different cases,
particularly as compared to the no-cavity case.
For the HH spectrum generated by a single irradiated Xe atom in a single cavity,
Fig.~\ref{fig:results_HHG_single_atoms}(b1) shows the generated HH spectrum for the case of no cavity,
and 
Figs.~\ref{fig:results_HHG_single_atoms}(b2) and \ref{fig:results_HHG_single_atoms}(b3) show the generated HH spectrum for, respectively, 
the cases of:  
(i) $\omega_{cav}$=6.45$\omega_{0}$ and $\varepsilon_{cav}$ = 0.229, 
corresponding to a side-harmonic shift of $\Delta M$=$\pm$1.0,
which means harmonic spacing of $\omega_{0}$ with
the generation of both odd and even harmonics;
and
(ii) $\omega_{cav}$=6.45$\omega_{0}$ and $\varepsilon_{cav}$ = 0.235, 
corresponding to a side-harmonic shift of $\Delta M$=$\pm$0.5.
Then, going beyond a single cavity,
Fig.~\ref{fig:results_HHG_single_atoms}(d1) presents results for the HH spectrum generated by a sequence of two cavities
with a single irradiated Xe atom in each.
Specifically here, these two cavities are the individual cavities of 
Figs.~\ref{fig:results_HHG_single_atoms}(b2) and (c2),
one corresponding to $\Delta M$=$\pm$1.0 while the other to $\Delta M$=$\pm$0.5,
which means harmonic spacing of $\omega_{0}$/2 
with the generation of odd and even harmonics as well as the middle harmonic between 
each odd-even pair.
Going to an even larger number of cavities, 
Fig.~\ref{fig:results_HHG_single_atoms}(d2) displays results for the HH spectrum generated by a sequence of ten cavities
with a single irradiated Xe atom in each.
All the ten cavities have a frequency of $\omega_{cav}$=6.45$\omega_{0}$,
while their $\varepsilon_{cav}$ spans the range of 0.238 to 0.256,
corresponding to shift magnitude $\left| \Delta M \right|$ of 
0.1 to 1.0 in a step of 0.1, 
which means harmonic spacing of $\omega_{0}$/10.
In general, there is also the possibility of blocking some harmonics in a generated HH spectrum to yield a corresponding modified spectrum.
As seen from all the above, in contrast to the HH spectrum generated without a cavity, the cavity-induced spectra include side harmonics around the odd harmonics, exhibit spectral intensities enhanced by orders of magnitude, and extend the plateau and cut-off regions to much higher harmonic orders.
Overall, the demonstrated cavity control over the generated HH spectrum is extremely rich.

Since one of the prominent uses of HHG is the production of attosecond pulses~\cite{paper4_attosec_pulse_gen_exp_2001},
Fig.~\ref{fig:results_HHG_single_atoms} also presents results for the 
attosecond pulse sequence that can be generated in the different cases.
Fig.~\ref{fig:results_HHG_single_atoms}(c1) shows the temporal attosecond pulse sequence generated 
from the cutoff region (beyond the 26th harmonic, H26) 
of the no-cavity HH spectrum presented in 
Fig.~\ref{fig:results_HHG_single_atoms}(b1),
after setting all the HH spectral phase to zero,
i.e., it is the corresponding transform-limited (TL) attosecond pulse sequence.
As is well known, and seen, the generated attosecond pulses are spaced by $T_{0}$/2,
with $T_{0}$ being the optical period of the strong cw field irradiating the Xe atom.
Fig.~\ref{fig:results_HHG_single_atoms}(c2) shows the TL attosecond pulse sequence that generated from the cutoff region (beyond H26) of the cavity-induced HH spectrum presented in 
Fig.~\ref{fig:results_HHG_single_atoms}(b2).
Since the harmonic spacing in this spectrum is $\omega_{0}$, 
as seen, the generated attosecond pulses are spaced here by $T_{0}$,
which is impossible impossible without the cavity
Similarly, Fig.~\ref{fig:results_HHG_single_atoms}(c3) shows the TL attosecond pulse sequence generated from the cutoff region (beyond H26) of the cavity-induced HH spectrum presented in panel (b3).
Here, the generated attosecond pulse sequence is composed of a pulse every 2$T_{0}$ 
accompanied by two weaker sub-pulses that are shifted from it by about $\pm$0.5$T_{0}$. 
To further illustrate the richness and control potential also in the production of the attosecond pulse sequence, 
Fig.~\ref{fig:results_HHG_single_atoms}(c4) shows the TL attosecond pulse sequence 
generated from the cutoff region (beyond H26) of the cavity-induced HH spectrum presented in Fig.~\ref{fig:results_HHG_single_atoms}(b3), 
after the odd harmonics are blocked, 
i.e., the spectrum contains here only the 
side harmonics with $\Delta M$=$\pm$0.5 around each odd harmonic. 
This brings the HH spectrum to the case of harmonic spacing of $\omega_{0}$, 
with the existing harmonic orders 
shifted by $\omega_{0}$/2 as compared to the HH spectrum of 
Fig.~\ref{fig:results_HHG_single_atoms}(b2).
Following this harmonic spacing of $\omega_{0}$, 
the TL attosecond pulse sequence that is generated here has a temporal spacing of $T_{0}$,
similar to the case pulse sequence presented in Fig.~\ref{fig:results_HHG_single_atoms}(c2).


In summary, using non-Hermotian Floquet theory, we
demonstrated cavity-controlled HHG, 
where HHG by a cw classical field irradiating a ground-state atom 
is controllable by placing the irradiated atom inside a single-mode 
cavity that is initiated even with a single photon. 
The control is based on cavity coupling of 
cavity-free atomic Floquet to form polaritonic Floquet states that their population allows to generate side harmonics around the standard no-cavity odd harmonics, which are the only ones generated 
without a cavity.
The different possible cavity-
controlled HH spectra, including also the ones resulting from several cavities in a row, enable the production of 
attosecond-pulse sequences that are different from
the one produced without a cavity. 
The results and degree of cavity control are extremely sensitive to the 
different characteristics of the system, 
including the irradiated atom, the strong cw field and the cavity,
hence their judicious selection is highly required.
In this spirit, the new theoretical framework established here opens avenues for further cavity control of the HHG process as well as other strong-field photo-processes.



\bibliography{References}




%
%
%


\end{document}